\RequirePackage{fix-cm}
\documentclass[twocolumn]{svjour3}          % twocolumn
\smartqed  % flush right qed marks, e.g. at end of proof
\usepackage{import}
\usepackage{graphicx}
\usepackage{url}
\usepackage{enumitem}
\usepackage{balance}
\journalname{09-03-2018; Working Paper;}
\begin{document}

\title{Review of Blockchain Technology and its Expectations: Case of the Energy Sector \thanks{This work has been funded by the UK EPSRC funded Refactoring Energy Systems (EP/R007373/1) and Household-Supplier Energy Market (EP/P031838/1) projects.}
}


\author{Ruzanna Chitchyan         \and
        Jordan Murkin %etc.
}


\institute{R. Chitchyan \and J. Murkin \at
              Department of Computer Science, University of Bristol\\ Merchant Venturers Building, Woodland Road, Bristol, BS8 1UB, UK\\
             % Tel.: +123-45-678910\\
   %           Fax: +123-45-678910\\
              \email{r.chitchyan@bristol.ac.uk} \\
              \email{jordan.murkin@bristol.ac.uk}
}

\date{Received: date / Accepted: date}

\maketitle

\begin{abstract}
This article suggests that the worldwide relevance of blockchain technology is motivated by the changes that it is expected to cause in: (i) the way that business is organised and (ii) regulated, as well as (iii) by the way that it changes the role of individuals within a society. The article presents an overview of the features of blockchain technology. It then takes a closer look into the developments within the energy sector across the world to gain a preliminary indication of whether the stated expectations are coming to reality. As a result of this review, we remain cautiously optimistic that blockchain technology could deliver the expected impact. 

\keywords{blockchain \and distributed ledger technology \and energy sector\and peer-to-peer energy trading}
\end{abstract}

\section{Introduction} \label{intro} 

In recent years academia and industry alike have been excited about blockchain. Blockchain has been proclaimed to be the next biggest technological breakthrough since the invention of internet. It is expected to revolutionise not only the technical structure of our communication and information technology, but also the very fabric of societies. The list of the changes expected to come through the blockchain are many, including: %`from replacing the technical stack of the internet itself, to changing the way that business is conducted and regulated, and even revolutionising the current consumerist society. Yet, the essence of all changes these can be condensed to the notions of secure desentralisation 

\textit{Changing the way business is conducted.} 
Until now societies needed trusted intermediaries to mediate most types of business transactions: for instance, individuals entrust their savings to a bank for safekeeping and interest accumulation and the bank loans these savings out to other individuals at higher interest rate; farmers deliver their produce to supermarkets who re-sell these to consumers at higher prices; energy generators sell their outputs to suppliers who re-sell the energy to end users. In all these cases both producers and consumers know and trust the intermediary (e.g., the bank has good reputation, farmers know this supermarket which is likely to work with other farmers around the given area, the energy distributor has “green” credentials) but do not know or trust each other. As blockchain provides cryptographic trust through technology design, whereby anonymous parties can transact without the possibility of cheating, intermediaries will no longer be necessary beyond the technical platform provision. This phenomenon has already commenced, to some degree, with the internet whereby virtual organisations (such as Airbnb and Uber) deliver the platform for individuals to transact with each other. Yet, presently, these organisations are monopolising platform delivery and still imposing substantial intermediation costs. As blockchain gets integrated into the ICT infrastructure, such monopolisation would become impossible.
% * <david.ferguson@edfenergy.com> 2018-02-16T22:06:01.589Z:
% 
% I think there's also the case that consumers have, by default, trust in 'the system'
% 
% ^.
% * <david.ferguson@edfenergy.com> 2018-02-16T22:00:00.990Z:
% 
% energy generators sell to suppliers (not network operators]
% 
% ^.

\textit{Changing the way business is regulated.} Until now societies have required regulators to ensure that businesses operate within legal frameworks: for instance, land registry authorities are to assure correct record keeping for land ownership, financial auditors are to assure proper fund handling and absence of embezzlement, competition authorities to oversee fair pricing and so on. As blockchain, along with its smart contracts, provides transaction record transparency, as well as imposing rules defined within contracts upon all transactions, regulatory and legal compliance checks become a prerequisite for any transaction completion. 

\textit{Changes the role of individuals within society.} Today we view ourselves as ``consumer societies", where individuals are generally passive consumers. However, the peer-to-peer transactive nature of blockchain encourages individuals into both productive and consumptive roles. The individuals are no longer passive consumers, but are active prosumers (i.e., \textit{pro}ducers and con\textit{sumers}). This again, is an ongoing process already today, with such examples as music production crowdsourcing by individual artists \cite{PledgeMusic}, microlending by peers \cite{Zidisha}, peer-to-peer file sharing, and micro-generation in the energy sector \cite{brooklynMicrogrid}. Common adoption of blockchain platforms would transform such activities from niche to norm. 

%\textit{Changing the technical structure of the internet}: xxx
%
As discussed, a diverse set of changes are expected across all walks of society, all of which are driven by the \textbf{\textit{secure decentralisation of social and technical structures}} enabled through blockchain technology.

In this paper we first present the makeup of blockchain technology, discussing what exactly distinguishes it and how (sections \ref{tech_rev} and \ref{bcPlatform}). We then review the academic and industrial state of the art with respect to blockchain technology for a specific industry - the energy sector - to analyse how blockchain technology has affected it (section \ref{energySec}). We finally consider how the present findings stand up against the stated expectations, and if there are any indications of realisability of these expectations (section \ref{conclusion}). As a result of this exercise, we remain cautiously optimistic that the blockchain technology could deliver the expected impact. 
\section{Technology Review} \label{tech_rev}
 
\subsection{The makeup of the blockchain technology}

A blockchain is a distributed database in which transaction records are collected into groups - called blocks - and stored with a reference to the previous block, forming an ever growing chain of blocks. These blocks are created by members of the network, known as \textit{miners}, who validate the transactions and are rewarded for their contribution.

Blockchains provide two characteristics that make them attractive as a transaction recording solution:

\begin{enumerate}
	\item \textbf{Cryptographic immutability and verifiability} making it quite impossible to modify transaction records once committed, thus ensuring secure transactions;
    \item \textbf{Distributed consensus} allowing anonymous individuals across a peer-to-peer network to come to agreement on the state of the network, thus removing the need of a centralised agreement mediator/ organisation.
\end{enumerate}

\subsubsection{Cryptographic immutability and verifiability}
Blockchains provide immutability and verifiability by pairing two existing technologies: hash functions (such as SHA256 \cite{damgaard1989design}) and Merkle trees \cite{merkel1988}.

A \textit{hash function} is an algorithm that takes arbitrary data and outputs a fixed-sized bit string known as a hash. Hush functions are one-way mapping functions, i.e., given the output of the function, there is no way of reverse engineering the input that generated the given bit-string\footnote{Except through ``brute force": searching through (the infinite) set of possible inputs, hashing them, and comparing the results, in the hope that (at some point) a match would be found. This, however, is an infinitely expansive process in terms of time and computational resources.}.
They are also quick to compute and deterministic, i.e., given the same input, the function will always produce the same output. Moreover, a small change in the input will dramatically change the output. In a blockchain, a hash function is used to represent the data content of a block with a fixed-length bit-string.

\begin{figure}[!ht]
\caption{A simple Merkle tree}
\label{FigMerkeling}
  \centering
  \includegraphics[scale=0.35]{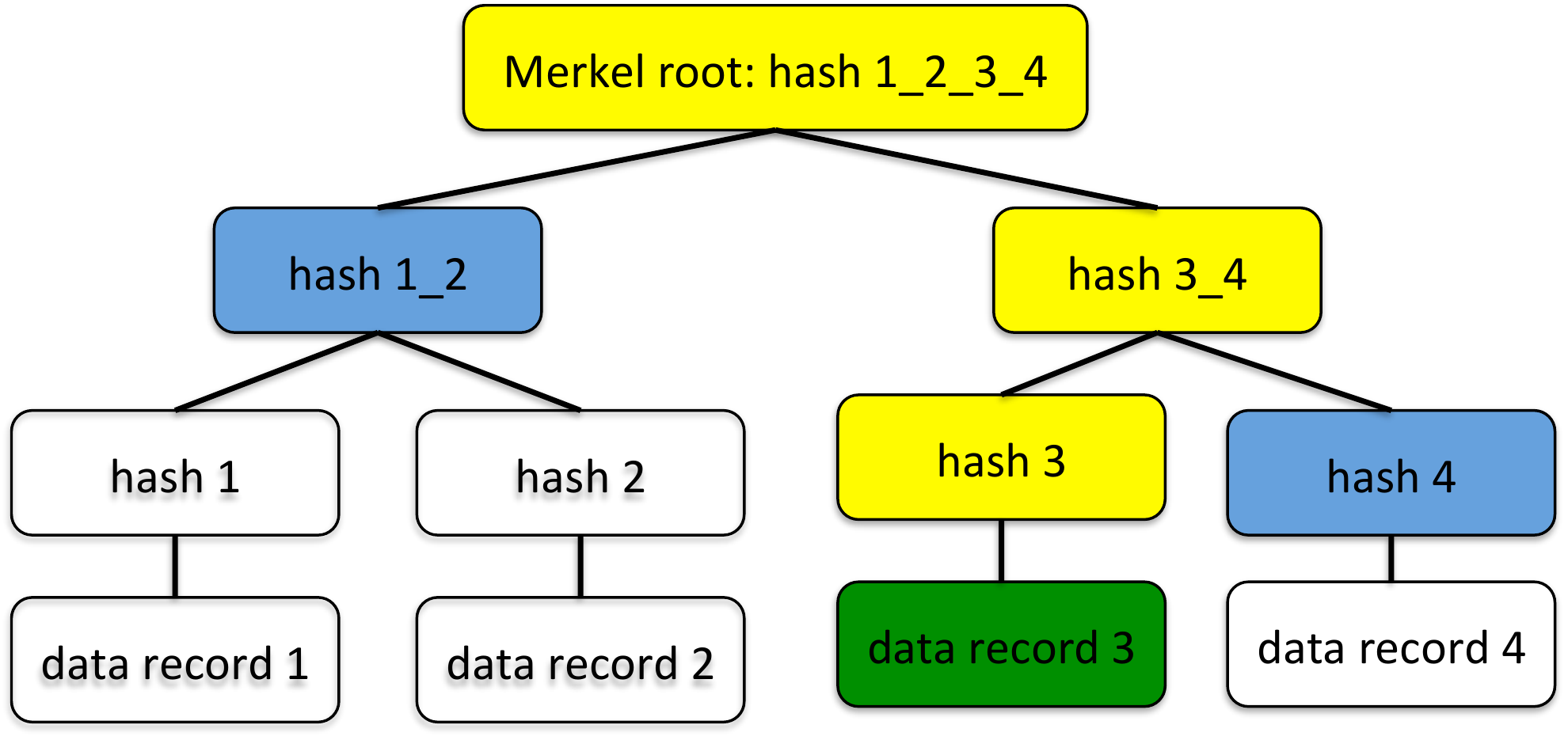}
\end{figure}

\textit{Merkle trees} are then used to structure the records in blocks, and support efficient verification of the chain's data authenticity. In a simple Merkle tree all data is contained in leaf nodes.
Each record is then hashed, and the parent nodes are layered upon the leaves by combining pairs of hashes of the lower level nodes and calculating a new hash node, as shown in Fig. \ref{FigMerkeling}.

Each block of a blockchain contains the top level hash of this tree, known as the \textit{Merkle root}, along with all the transactions that the block includes. Each block also contains the record of the hash of the previous block in the chain. When the hash of a new block is calculated, the new block becomes inseparably connected to its predecessor: should any part of the records in any of the recorded blocks change, the whole blockchain from that point onwards will diverge from its previous version.

Thus, use of Merkle trees ensure that no change can go unnoticed to previously agreed upon and recorded transactions. They also provide easy verification on the presence of a given record within the chain \cite{merkelingButerin2015} using the so called \textit{Merkle proof}. Given a top level hash (e.g., the Merkle root in Fig.\ref{FigMerkeling}), the node of data (e.g., \emph{data record 3} in Fig.\ref{FigMerkeling}), and the set of hashes that are used to integrate the current data with the rest of the tree (e.g., \emph{hash1\_2} and \emph{hash 3} in Fig.\ref{FigMerkeling}), a Merkle proof allows any interested party to verify that the given data (e.g., \emph{data record 3} in Fig.\ref{FigMerkeling}) is correctly and uniquely integrated into the chain without the party requiring knowledge of every individual child node (e.g., by comparing the newly calculated Merkle root for the given data record with the given one).

\subsubsection{Distributed consensus}
Distributed consensus is the process of coming to an agreement between individuals distributed across a peer-to-peer network. In a blockchain this is required for agreeing that a block validated by some miner should be added to the chain. Since miners are rewarded when their block is added to the chain, they would compete against each other and a number of candidate blocks could be available at any given time. The agreement is achieved through distributed consensus algorithms (DCA) \cite{ConsensusMechanisms}. Each blockchain platform uses its own flavour of DCA. Some examples are: 

\begin{enumerate}
\item \textit{Proof of Work} (PoW) \cite{nakamoto2009,back2002hashcash}, which operates by setting a target value, which must not be exceeded by the value of the hash for a given block, if that block is to be acceptable for addition to the blockchain (e.g., in Bitcoin \cite{bitcoin} and Ethereum \cite{ethereum}). This target is adjusted so that on average one node in the network will find a block with such a value within a given time interval (e.g, every 10 min. in Bitcoin network). The node that finds such a block can add this to the chain. This creates competition between miners to be the first to find an acceptable hash value.
\item \textit{Proof of Stake} (PoS) \cite{tendermint,casper2017} works by choosing the node that will be able to form the next block on basis of random selection from amongst the nodes that have maintained unspent currency within the blockchain network for the longest period of time (e.g., in Peercoin \cite{peercoin}) or of the largest size deposit (e.g., BlackCoin \cite{blackcoin}).

\item Proof of Burn (PoB) \cite{proofOfBurn} (e.g., in Slimcoin \cite{slimcoin}) where the next block creating node is chosen from amongst those who demonstrate “burning” some of their coins by sending these to a verifiably unspendable address.
\end{enumerate}

Whichever flavour of DCA is used, they all are structured to incur a demonstrable cost to the mining nodes.
Those who complete a block formation task are rewarded with a payment comprised from the fees that each blockchain participant pays to ensure completion of their transactions and a preconfigured amount of the network's cryptocurrency. 
Thus, the consensus mechanism aims to ensure that for a given node it is very expensive to attack the network and more profitable to help maintain it.  

Blockchains are maintained and operated by the blockchain network - an open membership peer-to-peer network of computers which redundantly store the data logs. This redundancy ensures that the network has no single point of failure or target for attacks. But it also necessitates a process by which all nodes on the network are able to ensure a consistent state of the blockchain copies \cite{aliPhD2017}. This process too is furnished though the above discussed DCA, by which all nodes on the network (or at least the clear majority, i.e., 51\% of them) accept that the longest blockchain (i.e, the version with the largest number of signed blocks) is the valid one. As noted above, the DCA imposes a cost upon miners for signing blocks, thus the longest chain is also the one which has the largest accumulated cost. Should a malicious node wish to alter a transaction committed to some previous block, it must recreate the block to-be-changed, as well as all the subsequent blocks (else the change will be immediately marked as invalid due to broken hashes). Such a malicious node would thus need to incur the cumulative cost for change, including also the cost of generation of the newest block, to become the current longest chain and have this accepted by the network.
This, however, is prohibitively expensive in terms of processing power and the costs imposed by a given DCA that a node would have to amass.

In summary, blockchains are distributed, decentralised, multi-access databases with cryptographic security of data records. Yet, the very nature of these databases requires a peculiar operating environment, such as dedicated network of peers to support it and a cost to committing transactions. So when are these databases to be preferred over the more traditional ones? 

\subsection{To blockchain or not to blockchain?}
As any technology, blockchain too has a specific set of use cases where it is particularly well suited. It is generally recognised \cite{Mattila2016,Greenspan2015,dltUKGov2016} that to be particularly well suited for the blockchain, a business case should require:

\begin{enumerate}
	\setlength\itemsep{0.7em}
    
	\item Use of a database, as the basic purpose of the blockchain is still to order and record transactions;
    \item This database must be shared amongst multiple users wishing to write to it to commit their own transactions;
    \item The transactions are interdependent, i.e., the order of the transactions matters (e.g., the investor must pay money before the borrower pays interest on it);
    \item The writers do not trust each other as they may have conflicting interests (e.g., investors may wish to gain interest without paying, while borrowers may wish to get money without paying interest); or simply have no sufficient information about each other;
    \item There is a need for disintermediation, i.e., when no third party (such as a bank for investment and borrowing) is suited to act as a trusted intermediary for all writers for one reason or another (e.g., cost and/or speed of intermediation under micro-lending schemes \cite{Zidisha}, ideology, etc.). 
\end{enumerate}

\subsection{And if to blockchain, will it scale?}
% https://www2.deloitte.com/content/dam/Deloitte/uk/Documents/Innovation/deloitte-uk-blockchain-full-report.pdf
% https://blockchainhub.net/blockchains-and-distributed-ledger-technologies-in-general/
%
Securing decentralised transactions was the primary goal of blockchain technology, which is now sufficiently resolved. However, the mass adoption of this technology requires a resolution to one other challenge - that of scaling up the volume and speed of transaction processing over a blockchain infrastructure. There presently is an unresolved tension between scalability, security, and decentralisation concerns \cite{dahlquist2017}, as only two of them at a time can (so far) be addressed satisfactorily with blockchain technologies. The established PoW DCA, for instance, is extremely secure and fully decentralised, as the computational resources (in terms of processing power and expended energy) required to falsify the records on the accepted chain are completely prohibitive. Yet, the very costs required to guarantee security (e.g., resources and time expended on hashing, consensus, and competition between miners) inhibit its scalability. 

%To address the tension within the scalability, security, and decentralisation trilemma, a number of solutions are proposed that relax the requirements that each miner must keep a copy of the full transaction history and validate the full chain to confirm transaction legitimacy. Such relaxations come with their own tradeoffs.

To address the tension within the scalability, security, and decentralisation trilemma, a number of solutions have been proposed that relax the requirements of the blockchain mechanisms. Such relaxations come with their own tradeoffs.

% Scalability and verification
% [35]Jason Teutsch and Christian Reitwießner. “A scalable verification solution for blockchains”. 2017.
% [36] Guy Zyskind. “Efficient Secure Computation Enabled by Blockchain Technology”.
% Master Thesis. Massachusetts Insitute of Technology, 2016.
%Trust and verifiable computation for smart contracts in permissionless blockchains by DOMINIK HARZ
%
\subsubsection{Permissioning}
To start with, blockchain began as an open, so-called \textit{permissionless} network, where any interested entity can act as a miner. Yet presently blockchains are used with varying degree of \textit{permissioning}, which helps to reduce the competition between miners and latency of consensus at the expense of decentralisation and security.

\textit{Public blockchains}
are fully open to everyone; anyone may inspect and participate in the network, viewing transactions, creating transactions and mining blocks. The original blockchain concept was a public blockchain, and this was important to provide the transparency and immutability required for trustless, intermediary-free, decentralised transactions to take place.

\textit{Consortium blockchains}
add a layer of permissioning, such that only selected organisations can validate transactions. Access to transaction history may also be restricted, but is in some cases kept public for full transparency. These blockchains are quicker and scale more easily, but at the cost of trust. Unlike public blockchains, the validators must be trusted by the users to validate transactions correctly.

Consortium chains are primarily used to increase automation of transactions between organisations, reduce cross-organisation transaction fees, improve standardisation, reduce fraud and increase auditability. For example, automatic settlement of trades between banking institutions could be handled by a blockchain using smart contracts where the validators in the network are the banks themselves. The Energy Web Foundation \cite{energyWeb} is already developing a consortium chain aiming to accelerate use of the blockchain infrastructure in the energy sector.

\textit{Private blockchains}, as the name suggests, are the most restrictive. These reserve validation task to a single organisation. Similarly to consortium chains, viewing of the transaction history may either be restricted or left public, depending on the use case. 

\subsubsection{Off Chain Transactions}
Moving some portion of transaction processing between account holders on a blockchain away from the chain itself is another currently popular direction for improved scalability.
This is done by using (a portion of) the assets currently present in the accounts of the transacting parties (say A and B) as fenced off payment guarantee for a specified time period (referred to as the \textit{funding transaction} \cite{lightningWhite}; lets say A and B each commit 10 coins for 1 week). The transacting parties can then undertake two way transactions within the agreed funding transaction limits (say A pays 5 coins to B, B pays 12 to A, then A pays three lots of 2 coins to B). At the end of the agreed time period the final account between participants is settled\footnote{This is a simplified explanation, in reality each individual transactions will also be cryptographically signed by both parties; parties will be able to redeem their funds before end of set time; penalties will be payable by non-cooperating parties for violating the transaction agreements \cite{lightningWhite}.} (e.g., now A has 10-5+12-2-2-2=11 coins and B has 10+5-12+2+2+2 = 9 coins) and recorded into the blockchain (a single summative transaction of B paying 1 coin to A instead of 5 actual transactions). 

This solution could be adapted to support not only bi-lateral, but also multilateral off-chain transactions, which guarantees fast transaction processing time for large volumes of translations without need for any intermediary, but for an account availability on the chain. It is also claimed to support privacy and anonymity of the transaction participants. Yet, the transparency of the whole transaction history between participants is lost, along with the trust and full accountability (e.g., for regulatory compliance and business practices).  

There are presently a few implementation efforts (e.g., Raiden \cite{raiden} and Lightning \cite{lightning} networks) on the way to integrate off-chain transaction handling into blockchain systems. The Lightning Bitcoin \cite{lightningBitcoing} fork of the Bitcoin blockchain for example, already runs with Lightning support. 

\subsubsection{Sharding}
\textit{Sharding} - the process of partitioning the blockchain into smaller sub-chains (termed \textit{shards}) is another direction of tackling the trilemma. Shards can be defined on basis of some specific criteria, such as account address space, use of specific applications, or geographical location, etc. With sharding \cite{sharding} most validator nodes would need to store and process transaction history data for specific shards only, not the full chain. Only a relatively small number of nodes would keep history of the full chain.

If validators are randomly assigned to the shard that they would validate for each block validation, and the result that the majority agrees upon is accepted and the current true state of the chain, the security of the transactions is still (mostly) assured, with improved throughput of the transaction processing.

Some of the main challenges to be address under sharding are, for instance: (i) slow cross-shard communication; (ii) complications in synchronisation of user requests that require atomic translation over two or more shards; (iii) risk of validator collusion in some consensus models where validators non-randomly choose what to validate (e.g., PoW), etc.

\subsection{Contractual Transactions}
The appeal of blockchain technology is magnified by the addition of \textit{smart contracts}\footnote{The name is somewhat misleading, as smart contracts do not (need to) have any real built-in intelligence.} to the blockchain infrastructure. 
A \textit{smart contract} \cite{Luu2016} is a piece of code (running inside a blockchain platform) that represents and enforces the protocol and any terms of a contract agreed upon by the contractual parties (e.g., seller and buyer of a product). The contracts can store arbitrary logic, such as restricting execution
%(i.e., without any intermediaries)
until a predefined criteria is met (e.g., a contract that only allows the sale of an item once a specific date/time has passed).
A smart contract is identified by an address with the contract's code housed within a blockchain. The contract is executed by sending transactions to its identifier address. 

The transactions executed via the contracts are secure (as these are recorded in a blockchain and enforced through the DCA), automated (and so are cheap and quick), and deterministic (always delivering the expected outcome, the code of the contract cannot be changed as it is also a part of the blockchain). All these properties foster dis-intermediation within businesses using such contracts (e.g., no need to have a broker to sell an item), and promise faster transactions at better price.

Yet, the smart contracts are also rigid and inflexible, which may cause difficulties (e.g., difficult to fix bugs identified in the code \cite{price2016}, or accommodate changes in context of the parties).

\section{Blockchain Platforms} \label{bcPlatform}
%https://docs.google.com/document/d/1RcVejjF5XPfqnQ55p2rCeDvZ_8602XsVgtHazQRJur8/edit#% 
A number of blockchain platforms are currently in development and use, each targeted for a specific use case.

\subsection{Bitcoin}
Bitcoin \cite{bitcoin} was the first to use blockchain and is subsequently the first cryptocurrency \cite{nakamoto2009}. It’s intention is to act as a decentralised electronic transaction system, in which individuals can store and transfer value between one another without the need for central authorities [1]. This value is represented by in Bitcoin tokens; with an issue limit of 21 million Bitcoins to provide scarcity of supply. Transactions on the network are paid for using Bitcoin. The platform operates under the PoW consensus mechanism and miners are rewarded in Bitcoin generation and through fees of the transactions they process. The transaction fees are specified by the transaction senders and transactions with the higher fees are prioritised by miners due to the increased profit from their processing. Thus users can reduce transaction fees for low priority transactions and vice versa. Each block size is limited to 1MB and the network intends to produce one block every 10 minutes, giving a maximum potential of \raisebox{-0.6ex}{\textasciitilde}7 transactions per second. Scalability improvement measures are under development, including increasing the block size (though this would reduce decentralisation as smaller miners would be unable to handle larger block mining) and integration of the Lightning network \cite{lightning} for off-chain processing.

\subsection{Ethereum}
Ethereum \cite{ethereum} was introduced in 2013 with the objective of providing a platform for decentralised applications. It took the concept of blockchain and incorporated a turing-complete scripting language with it. This allowed for applications themselves to be stored inside the blockchain where they can be used by anyone connected to the network.

The cryptocurrency in this network is Ether, which is used as a means to store and transfer value, as well as to pay for computation and transaction costs. Since Ethereum allows arbitrary code to be stored and executed inside the blockchain, there can be an infinite number of methods each requiring a different amount of computation and storage. To pay for this, Ethereum introduces the notion of \textit{gas}. Each opcode inside the Ethereum virtual machine is assigned a different amount of gas. Each transaction sent to the blockchain must specify: (i) a gas limit - the maximum computation the user is willing to incur; and (ii) a gas price - an amount of Ether the user will pay per gas unit. The gas price is used by the miners to prioritise transactions: those with the higher gas price are preferred by miners (similarly to Bitcoin). If the user does not specify enough gas for their requested transaction, the transaction will fail and, conversely, any excess gas provided will be refunded to the user.

Ethereum currently uses PoW protocol, but plans to move to a hybrid proof-of-work/proof-of-stake system \cite{casper2017} shortly and finally to a purely proof-of-stake mechanism in the future. The network can currently process \raisebox{-0.6ex}{\textasciitilde}15 transactions per second. Scalability measures are under discussion through planned integration with the Raiden network \cite{raiden} for off-chain processing, and efforts to implement sharding \cite{sharding}. 

% Add Casper reference: https://arxiv.org/pdf/1710.09437.pdf

\subsection{Ripple}
Ripple provides a real-time cross-border settlement and remittance network to banks, payment services providers, and corporates for the transfer of assets and money globally. The network is a permissioned blockchain with a consortium of approved validators, but the ledger is public. The network uses its own Ripple Protocol Consensus Algorithm \cite{rippleConsensus}. This is a vote-based consensus algorithm in which each validator has a vote and at least 80\% of voters must vote in favour for a transaction to be successful.

The network uses own currency (called XRP) for imposing anti-spam transaction fees and as a currency for value exchange.

\label{section:iota}
\subsection{Iota}
Iota \cite{iota} is not a blockchain, it is stated as a cryptocurrency for the IoT industry. Instead of using a blockchain, it is built upon the Tangle \cite{tangle}, a mechanism that ``succeeds the blockchain". The tangle does not have miners, instead opting for a user driven network; whenever a transaction is sent, the sender must authenticate two previous transactions. This allows the network to remain decentralised while also reducing transaction time and removing fees completely. As there are no miners, the computation required to run the network is significantly reduced, allowing nodes to run on devices with little computational power. This design choice also means that as more transactions are made, the network transaction time reduces further. 

\section{Blockchain in Energy Sector} \label{energySec}
Many - both in academia and industry - believe that the rise of blockchain could potentially foster innovative changes and facilitate the transition to the smart grid \cite{Mylrea2017}.
The concept of a decentralised electricity grid has been around for some time now \cite{Ipakchi2009}. Recently, the integration of energy storage devices as well as electric vehicles into the future electricity grid \cite{Farhangi2010} has initiated a wide discussion, as have studies on new control schemes for both energy storage and demand side response programmes \cite{Huang2012}. 

The idea of using blockchain in the energy sector is gaining an increasingly large interest.
%\subsection{Electric vehicles}
For instance, some researchers propose to integrate blockchain with electric vehicles (EV) \cite{Knirsch2017} so that EVs could use blockchain to find a nearby charging stations, while charging stations could bid for the opportunity to charge EVs. This mechanism would help find the best price and location for both EV users and charging stations, while at the same time providing privacy and security to the EVs.

%\subsection{Smart home / IoT}
Use of blockchain for IoT and, subsequently, energy efficiency in the smart homes is another area of active research  \cite{Dorri2017}. Here, blockchain could play a key role in data control and decision support for large scale IoT using smart contracts as means to communicate, automate and enforce rules between devices. As mentioned before (see \ref{section:iota}), a dedicated blockchain platform for IoT is already under development.

%\subsection{Emissions trading}
Blockchain was used to eliminate fraudulent behaviour in emissions trading \cite{Khaqqi2018}. Here an alternative Emissions Trading Scheme (ETS) backed by a blockchain was developed. Blockchain helped to ensure reliable, secure transactions and embed a reputation system to encourage investment into ETS in the long-term.

%\subsection{Guarantee of Origin certificates}
In a related work \cite{Castellanos2017} \cite{Imbault2017} the tokenisation of Guarantee of Origin (GoO) certificates \cite{GuaranteesofOriginOfgem} was considered. Since these certificates could act as a proof that a specified amount of electricity was generated through renewable sources, they could also form part of the emissions trading market. Thus, blockchain can support trade in such standardised certificates, as well as foster removal of the intermediaries from the market.

%\subsection{Peer-to-peer electricity trading}
Expanding on these certificate trading systems is the concept of peer-to-peer (P2P) electricity trading. With P2P trading systems the units of generated electricity themselves are recorded inside a blockchain, allowing the owner of this generation to market it to others. This enables energy generators and buyers (both large and small) to take ownership of their product, choices, and preferences, rather than solely rely on the grid as an intermediary \cite{Murkin2018}. Some researchers on p2p energy trading focused on the P2P energy market creation \cite{Kounelis2017} \cite{Murkin2018}, \cite{Hahn2017}, demonstrating that blockchain-based intermediary-free energy trading is possible and beneficial to the generators and buyers alike. Others studied the optimisation of energy resources \cite{Munsing2017} in P2P trading. In \cite{Mengelkamp2017}, the authors have developed a smart contract based closed double auction market mechanism wherein individuals could submit bids and sale orders for each market period and an automated contract would decide a market clearing price for the period.

A prototype trading platform was developed in \cite{Aitzhan2016} focusing on privacy and anonymity of users while also removing a single point of failure. This platform opted to remove a central price setting algorithm, instead providing users with encrypted channels to directly negotiate price between a buyer and a seller.

Scanergy \cite{Scanergy} \cite{MihaylovJARMABG15} research project is taking a different approach of incorporating blockchain into the electricity network. Here participants are provided with an incentive to export electricity and assist with the balancing supply and demand on the grid. For every kWh a household exports to the network that is consumed by another household, the exporter is credited with an NRGCoin \cite{MihaylovJMAN14} \cite{MihaylovJurado2014}. These NRGCoins can then be traded on a separate market like any other cryptocurrency. This system works using smart meter data and street-level substation data. The substation sees the total consumption and generation for the group of houses connected to it and smart meter data provides individual household information for the same period. Using the consumption and generation values from both the individual smart meters and the substation they are connected to, it is determined whether the exports of each house were consumed by another household. The producers are then credited with NRGCoins corresponding to the exported amounts. The smart meter data and substation data is also used to prevent tampering by ensuring the totals from the smart meters is the same as the totals from the substation.

Despite blockchain being a hot topic for energy researchers, the industry has taken the real lead in championing this technology (see Table \ref{TabIndustry}). 

\begin{table*}[!t]
\centering
\caption{Blockchain companies currently working in energy sector (adapted from \cite{15Firms})}
\label{TabIndustry}
\begin{tabular}{|p{2cm}|p{9.5cm}|p{4.5cm}|}
\hline
\textbf{Company}   & \textbf{Blockchain Use}  & \textbf{Operation at/since} \\ \hline
Alliander \cite{alliander}   & A smart energy company which has piloted a P2P energy trading platform. The energy production tokens (Juliets) are also exchangeable for goods and services within the piloting community. & The Netherlands, since 2017.  \\ \hline
Bankymoon \cite{bankymoon}  & A blockhain solutions company that introduced prepaid blockchain-enabled smart meters in Africa to help energy suppliers collect payment, as well as to enable humanitarian aid to be sent as energy via direct payments to smart meters at schools.& South Africa since 2015 \\ \hline
Conjoule \cite{conjoule} & Platform to support P2P trading among rooftop PV owners and interested public-sector or corporate buyers.                            & Germany: 2 pilots running since 2016 %Tokyo Electric Power Company (Japan) and Innogy (Germany)Op: Germany 
% * <rc256@le.ac.uk> 2017-12-13T02:23:09.072Z:
% 
% > Conjoule
% Renewables generators (e.g., homes) decide to whom to sell, community and ecosystem focus. 
% 
% ^ <rc256@le.ac.uk> 2017-12-13T02:26:16.121Z.
\\ 
\hline
Drift \cite{drift}& 
% * <rc256@le.ac.uk> 2017-12-13T02:28:57.496Z:
% 
% > Drift
% end user focused, supports price and renewable sources, represents end users.
% 
% ^.
Retail energy provider that uses blockchain, machine learning and high-frequency trading to provide better prices to customers and promote green energy use. & New York, USA since 2014\\
\hline
Greeneum \cite{greeneum}& 
% * <rc256@le.ac.uk> 2017-12-13T02:26:46.483Z:
% 
% certifies/enables green generation, but whole industry and prosumer focused.
% 
% ^.
P2P energy trading platform that incentivises renewable-based generation through GREEN tokens, global data collection and AI-based processing for energy industry optimisation. & 
Beta release in Cyprus, UK, Israel, Germany, Guinea, Argentina, US, India, Australia. Expects a product release by mid-2018\\ \hline
Grid+ \cite{gridplus}& 
% * <rc256@le.ac.uk> 2017-12-13T02:30:37.828Z:
% 
% > Grid
% retail provider, end user focused for price and source; also p2p trading between customers.
% 
% ^.
Retail provider (i.e.,buys on behalf of its customers at wholesale prices from outside) and P2P trading platform between Grid+ customers& 
%ConsenSys (NY USA)
Texas, USA since 2017
\\ \hline
Grid Singularity \cite{gridSingularity}  & 
Developing a blockchain-based core technology for the energy sector, focused on B2B provision; this technology is to underpin EWF& Internationally since 2016 (?)
\\ \hline
Electron \cite{electron}  & 
% * <rc256@le.ac.uk> 2017-12-13T02:33:05.395Z:
% 
% > Electron
% End user - price optimisation; p2p trading between end users;  Grid balancing is industry facing.
% 
% ^.
Automated energy supplier switching platform; also aims to to support P2P energy trading and grid-balancing& 
%U.K. startup with Siemens and National Grid, it won U.K. government support to scale up its platform.
UK, since 2016 
\\ \hline
Energy Web Foundation \cite{energyWeb}& Non-profit alliance between major energy players internationally, aimed at accelerating blockchain technology across energy market, and building an ecosystem around blockchain for energy. &
% * <rc256@le.ac.uk> 2017-12-13T02:34:28.947Z:
% 
% > Foundation
% Non-profit. Industry focused.
% * <rc256@le.ac.uk> 2017-12-13T02:35:04.895Z:
% 
% > ImpactPPA
% End-to-End provision with end user focus. Full industry replacement!?
% 
% ^.
%Manhattan Beach, California-based (USA)
Projects in preparation at Somaliland, Haiti, India, Argentina, since 2017\\ \hline
LO3 Energy (Exergy) \cite{LO3Energy}& 
P2P energy trading platform, aiming also at grid-level service provision (e.g., DER aggregation, balancing, wholesale trading). & NY, USA, since 2017 \\ \hline
MyBit \cite{myBit} & 
% * <rc256@le.ac.uk> 2017-12-13T04:05:18.258Z:
% 
% Focused on investment by individuals into hardware. Can be any hardware, including for renewables generation.
% 
% ^.
% * <rc256@le.ac.uk> 2017-12-13T04:05:14.184Z:
% 
% > MyBit
% p2p investment by small investors.
% 
% ^.
P2P investment into IoT hardware, such as connected solar panels. An investor can own a portion of tokenised hardware and get return per owned portion.& 
Alfa launch of platform in 2018\\ \hline
Ponton's \cite{ponton} Enerchain & A B2B solutions integration company that runs the Enerchain platform used for peer to peer blockchain-based energy trading at wholesale energy market by energy sector businesses. The traders anonymously send orders to a decentralised order book, which can also be used by other organisations. Enerchain does not require a central authority.    & 
The platform is used by some 30 European companies since 2016  \\ \hline
Power Ledger \cite{powerLedger} & 
P2P energy trading  (individuals and wholesale for utilities), EV charging,  transmission grid monitoring, P2P asset funding and asset ownership token trading. &  Australia and New Zealand, since 2016\\ \hline
SolarCoin Foundation \cite{solarCoing}  & The foundation aims to foster solar energy generation installations. It awards crypto-coins (for free, similar to air miles) to registered and verified solar energy producers. Each coin represents 1 MWh of produced solar energy. 
%SolarCoin comes on top of peer-to-peer energy platforms such as Power Ledger, Grid Singularity or LO3 Energy, and provides a solution to the whole solar value chain, utilities and governments. The coins are awarded to registered generators, and the value is created  top of government-backed subsidies.
    & Used in more than 13 countries since 2014  \\ \hline
Sun Exchange \cite{sunExchange} & P2P funding of solar PV installations in return for income on investment. Installations are for specific projects in Southern Africa to supply energy to schools, hospitals, and similar businesses. 
    & Southern Africa since 2015 \\ \hline
Veridium Labs \cite{veridium}   & 
Veridium is a financial technology firm aiming to create a new asset class that tokenizes natural capital. Each token will represent removal of 1 ton of greenhouse gases from the atmosphere, or equivalent natural capital preservation  activities (e.g., conserve 1 sq. meter of biodiverse tropical forest). Tokens will be issues for validated projects. This would be used by firms to conform with environmental impact mitigation regulations, and more generally embed environmental replacements into the cost of their products. & 
First project to commence in 2018  for Rimba Raya Biodiversity Reserve in Borneo, Indonesia, tokens will be traded internationally.\\ \hline
WePower \cite{wePower}  & A platform for P2P trading of renewable energy, as well as fund raising for renewable projects by pre-sale of energy to be generated in the future. 
& To commence in 2018 in Lithuania and Spain
\\ \hline
% Invirohub  \cite{invirohub}  & Invirohub is an integrated technology that incorporates smart meters telecoms infrastructure and proprietary software. 
% Aims to develop smart electricity, solutions giving users accurate detailed real time information.
%     & C3  \\ \hline

\end{tabular}
\end{table*}

There is a quickly growing number of startups as well as established energy players who are already tackling energy sector issues using blockchain technology.  Table \ref{TabIndustry} (expanded from \cite{15Firms}) presents a summary of some of the most prominent companies currently active in this space. 

As Table \ref{TabIndustry} demonstrates, there is a real breadth of the areas and purposes within energy sector where blockchain is being actively employed. Some, for instance, utilise blockchain for p2p energy trading aiming to eliminate retail intermediaries \cite{Brooklyn2016,BrooklynGrid,Exergy}; others, quite the opposite, use it for themselves becoming  more competitive and affordable energy retail intermediaries \cite{drift,gridplus}; some utilise it to support non-for-profit and charitable causes \cite{bankymoon}, while others base wholesale B2B energy trading solutions on it \cite{enerchain,gridSingularity}; some promote industry standardisation \cite{energyWeb}, others set out to monetise environmental resources and even their protection \cite{veridium}. 
%
%One of the best known projects is TransActive Grid \cite{Brooklyn2016} \cite{Exergy}, developed through a joint venture between LO3 Energy and ConsenSys. This project is active and working on a microgrid in Brooklyn, New York. Their system relies on replacing smart meters with their own custom meters, TransActive Grid (TAG) meters, to record data directly onto a blockchain. Recorded electricity on the network is tokenised and traded across an open decentralised market through their platform \cite{BrooklynGrid}.

Several observations (be it preliminary) surface from the review of the current blockchain-based businesses (summed up in Table \ref{TabIndustry}):
\begin{itemize}

\item Blockchain is used for \textit{starting up a new business structure and ecosystem} within the energy sector, competing against the incumbents. Thus, some businesses, like myBit \cite{myBit} foster investment into renewable generation hardware in such a way that, even if an individual is unable to purchase a whole device, he/she can invest into a portion of a tokenised hardware and get return per owned portion. Once generated, the energy production is recorded into a blockchain, over which software (acting as a wholesale or retailer trader) \cite{alliander,conjoule,electron,drift,gridplus} buys and sells energy. Software companies \cite{Brooklyn2016,gridSingularity,electron} (rather than the traditional distribution network operators and retailers) cater for the (increasingly more and more automated and optimised, e.g., with AI \cite{greeneum})  delivery of and accounting for the energy generation and consumption. To facilitate the emergence and operation of this new ecosystem, a standardisation effort is already under way (e.g.,\cite{energyWeb}).

\item Driven by this newly emerging ecosystem, \textit{new kinds of energy services} are also beginning to emerge (e.g., balancing the supply and demand within the grid or remotely scheduling consumer device operation, such as EVs charging, running washing machines, etc., in response to the generated every availability) \cite{LO3Energy}.

\item \textit{Renewable forms of energy are very much the core} of the products delivered via a blockchain infrastructure. Some businesses foster trading of the generated renewable energy \cite{alliander,conjoule,Brooklyn2016,drift,greeneum}, while others support adoption \cite{myBit,sunExchange} and better utilisation \cite{gridplus} of the household-level generation assets;

\item Many of these businesses are \textit{end-user focused}, aiming to obtain better energy prices for the end users \cite{alliander,conjoule,electron,drift,gridplus}, or access to/participation in the energy generation endeavour \cite{myBit,sunExchange};

\item Finally, the incumbent energy businesses have also identified blockchain as a potent technology, and have started using it for business process optimisation and inter-business communication \cite{energyWeb,gridSingularity}.
\end{itemize}
\section{Concluding Thoughts} \label{conclusion}
This article presents an overview of the basic features of blockchain technology. 
In the introduction to this article we suggested that the high relevance of blockchain technology is motivated by the changes that it is expected to cause in: (i) the way that business is organised and (ii) regulated, as well as (iii) by the way that it changes the role of individual within a society. We then took a closer look into the developments within the energy sector (see section \ref{energySec}) across the world, to gain a preliminary indication of the effects that blockchain has caused within this sector and assess whether the current trends within energy sector confirm or refute the stated expectations.

As noted above (section \ref{energySec}), blockchain has indeed initiated a change of business ecosystem and organisation within the energy sector. For instance, by
\begin{itemize}
\item removing the need for intermediaries (currently often played by the established energy utility providers) through direct p2p energy trading and even allowing wholesale generators to trade directly with the end users/consumers (e.g., \cite{alliander,conjoule,electron,gridplus});
\item enabling individuals and communities to initiate energy production projects that require expensive investment which they could access through crowd-funding (e.g., \cite{myBit,sunExchange});
\item transposing a number of energy services (e.g., billing, supplier switching) form established retailers and utilities to software platform providing companies which automate these services and deliver them at a fraction of the ``normal" costs;
\item allowing for a number of new, previously non-existent services to emerge, e.g., to optimise the price of a household's energy consumption by scheduling such tasks as washing and vehicle battery charging; or balancing energy grid.
\end{itemize}
  
We also observe a change of the role that the households play in energy sector: these are not simply consumers, but active owners and/or investors into energy hardware and software infrastructure, generators and sellers of energy, as well as buyers and consumers. In the past intermediated market, the capacity of an individual household would be considered too small for acting as either a viable energy seller or investor. Now, with no intermediaries and a low barrier to market entry, households and individuals not only can assume new roles, but are also actively encouraged to do so by the new businesses that rely on active household investors and prosumers. 

Regulation of businesses through blockchain technology in energy sector is currently somewhat under-developed. This has largely to do with blockchain-based energy businesses being a relatively novel phenomena. However, recent developments worldwide indicate that regulators recognise the potential of this technology in both enabling more efficient energy systems, and facilitating their regulation. For instance, a law enabling self-consumption by individuals and small communities has already been passed in France \cite{frenchLaw2017} and there is regulations support for (limited) P2P energy trading in Germany, Netherlands, and USA \cite{australianReport2017}.
Furthermore, most businesses discussed in Table \ref{TabIndustry} (e.g., \cite{electron,LO3Energy,powerLedger}) already use smart contracts over blockchain for specifying and enforcing energy trading contracts to support efficient trading, as well as business transparency and trustworthiness. 

In summary, looking at the trends within the energy sector, we suggest that the impacts expected from blockchain technology have already started to manifest in practice, and seems to be grounded in reality, though there still is a long way to go for its full-scale rooting with the energy sector. Thus, we remain cautiously optimistic that these expectations will be realised in not-so-distant future. 
\balance

\bibliographystyle{spphys}       % APS-like style for physics
\bibliography{jisa_2017}  

\begin{thebibliography}{10}
\providecommand{\url}[1]{{#1}}
\providecommand{\urlprefix}{URL }
\expandafter\ifx\csname urlstyle\endcsname\relax
  \providecommand{\doi}[1]{DOI \discretionary{}{}{}#1}\else
  \providecommand{\doi}{DOI \discretionary{}{}{}\begingroup
  \urlstyle{rm}\Url}\fi

\bibitem{PledgeMusic}
Pledgemusic.
\newblock \urlprefix\url{https://www.pledgemusic.com/}

\bibitem{Zidisha}
Zidisha.
\newblock \urlprefix\url{https://www.zidisha.org/}

\bibitem{brooklynMicrogrid}
Brooklyn microgrid.
\newblock \urlprefix\url{http://brooklynmicrogrid.com/}

\bibitem{damgaard1989design}
I.B. Damg{\aa}rd, in \emph{Conference on the Theory and Application of
  Cryptology} (Springer, 1989), pp. 416--427

\bibitem{merkel1988}
R.C. Merkel, in \emph{Advances in Cryptology}, \emph{Lecture Notes in Computer
  Science}, vol. 293, ed. by C.~Pomerance (Springer, 1988), \emph{Lecture Notes
  in Computer Science}, vol. 293

\bibitem{merkelingButerin2015}
V.~Buterin.
\newblock Merkling in ethereum (2015).
\newblock
  \urlprefix\url{https://blog.ethereum.org/2015/11/15/merkling-in-ethereum/}

\bibitem{ConsensusMechanisms}
Waves.
\newblock Review of blockchain consensus mechanisms.
\newblock
  \urlprefix\url{https://blog.wavesplatform.com/review-of-blockchain-consensus-mechanisms-f575afae38f2}

\bibitem{nakamoto2009}
S.~Nakamoto.
\newblock Bitcoin: A peer-to-peer electronic cash system (2009).
\newblock \urlprefix\url{https://bitcoin.org/bitcoin.pdf}

\bibitem{back2002hashcash}
A.~Back, et~al.,   (2002)

\bibitem{bitcoin}
Bitcoin.
\newblock \urlprefix\url{https://bitcoin.org/}

\bibitem{ethereum}
Ethereum.
\newblock \urlprefix\url{https://www.ethereum.org/}

\bibitem{tendermint}
J.~Kwon.
\newblock Tendermint: Consensus without mining (2014).
\newblock \urlprefix\url{https://tendermint.com/static/docs/tendermint.pdf}

\bibitem{casper2017}
V.~Buterin, V.~Griffith, CoRR \textbf{abs/1710.09437} (2017).
\newblock \urlprefix\url{http://arxiv.org/abs/1710.09437}

\bibitem{peercoin}
S.~King, S.~Nadal.
\newblock Ppcoin: Peer-to-peer crypto-currency with proof-of-stake (2012).
\newblock \urlprefix\url{https:// decred.org/research/king2012.pdf.
  https://web.archive.org/save/https://decred.org/ research/king2012.pdf}

\bibitem{blackcoin}
P.~Vasin.
\newblock Blackcoin’s proof-of-stake protocol v2.
\newblock \urlprefix\url{URL http://blackcoin.co/
  blackcoin-pos-protocol-v2-whitepaper.pdf}

\bibitem{proofOfBurn}
Bintcoinwiki.
\newblock Proof of burn.
\newblock \urlprefix\url{https://en.bitcoin.it/wiki/Proof_of_burn}

\bibitem{slimcoin}
Slimcoin. a cryptocurrency for the long term.
\newblock \urlprefix\url{http://slimco.in/}

\bibitem{aliPhD2017}
M.~Ali, Trust to trust design of a new internet.
\newblock Ph.D. thesis, University of Prinston (2017).
\newblock Blockchainbased internet with storage off chain and saclability

\bibitem{Mattila2016}
J.~Mattila, The blockchain phenomenon -- the disruptive potential of
  distributed consensus architectures.
\newblock Tech. Rep. Working Paper No. 38, ETLA (2016).
\newblock \urlprefix\url{https://www.etla.fi/category/julkaisut/}

\bibitem{Greenspan2015}
G.~Greenspan.
\newblock Avoiding the pointless blockchain project (2015).
\newblock
  \urlprefix\url{https://www.multichain.com/blog/2015/11/avoiding-pointless-blockchain-project/}

\bibitem{dltUKGov2016}
M.~Hancock, E.~Vaizey, Distributed ledger technology: beyond block chain.
\newblock Tech. rep., UK Government Office for Science, GOV.UK (2016)

\bibitem{dahlquist2017}
O.~Dahlquist, L.~Hagstrom, Scaling blockchain for the energy sector.
\newblock Master's thesis, University of Uppsala (2017).
\newblock Scalability, decentralisation, security trilemma and energy sector

\bibitem{energyWeb}
Energy web foundation.
\newblock \urlprefix\url{http://energyweb.org/}

\bibitem{lightningWhite}
J.~Poon, T.~Dryja.
\newblock The bitcoin lightning network: Scalable off-chain instant payments
  (2016).
\newblock \urlprefix\url{https://lightning.network/lightning-network-paper.pdf}

\bibitem{raiden}
Raiden network.
\newblock \urlprefix\url{https://raiden.network/}

\bibitem{lightning}
Lightning network.
\newblock \urlprefix\url{http://lbtc.io/}

\bibitem{lightningBitcoing}
Lightningbitcoin.
\newblock \urlprefix\url{http://lbtc.io/}

\bibitem{sharding}
A.~Zamyatin.
\newblock On sharding blockchains (2017).
\newblock
  \urlprefix\url{https://github.com/ethereum/wiki/wiki/Sharding-FAQ#on-sharding-blockchains}

\bibitem{Luu2016}
L.~Luu, D.H. Chu, H.~Olickel, P.~Saxena, A.~Hobor, in \emph{Proceedings of the
  2016 ACM SIGSAC Conference on Computer and Communications Security} (ACM, New
  York, NY, USA, 2016), CCS '16, pp. 254--269.
\newblock \doi{10.1145/2976749.2978309}.
\newblock \urlprefix\url{http://doi.acm.org/10.1145/2976749.2978309}

\bibitem{price2016}
R.~Price, Business Insider  (2016)

\bibitem{rippleConsensus}
N.Y. David~Schwartz, A.~Britto,   (2014).
\newblock
  \urlprefix\url{https://ripple.com/files/ripple_consensus_whitepaper.pdf}

\bibitem{iota}
Iota.
\newblock \urlprefix\url{https://iota.org/}

\bibitem{tangle}
S.~Popov.
\newblock The tangle (2017).
\newblock \urlprefix\url{https://iota.org/IOTA_Whitepaper.pdf}

\bibitem{Mylrea2017}
M.~Mylrea, S.N.G. Gourisetti, in \emph{2017 Resilience Week (RWS)} (2017), pp.
  18--23.
\newblock \doi{10.1109/RWEEK.2017.8088642}

\bibitem{Ipakchi2009}
A.~Ipakchi, F.~Albuyeh, IEEE Power and Energy Magazine \textbf{7}(2), 52
  (2009).
\newblock \doi{10.1109/MPE.2008.931384}

\bibitem{Farhangi2010}
H.~Farhangi, IEEE Power and Energy Magazine \textbf{8}(1), 18 (2010).
\newblock \doi{10.1109/MPE.2009.934876}

\bibitem{Huang2012}
L.~Huang, J.~Walrand, K.~Ramchandran, in \emph{2012 IEEE Third International
  Conference on Smart Grid Communications (SmartGridComm)} (2012), pp. 61--66.
\newblock \doi{10.1109/SmartGridComm.2012.6485960}

\bibitem{Knirsch2017}
F.~Knirsch, A.~Unterweger, D.~Engel, Computer Science - Research and
  Development  (2017)

\bibitem{Dorri2017}
A.~Dorri, S.S. Kanhere, R.~Jurdak, P.~Gauravaram, in \emph{2017 IEEE
  International Conference on Pervasive Computing and Communications Workshops
  (PerCom Workshops)} (2017), pp. 618--623.
\newblock \doi{10.1109/PERCOMW.2017.7917634}

\bibitem{Khaqqi2018}
K.N. Khaqqi, J.J. Sikorski, K.~Hadinoto, M.~Kraft, Applied Energy \textbf{209},
  8  (2018).
\newblock \doi{https://doi.org/10.1016/j.apenergy.2017.10.070}.
\newblock
  \urlprefix\url{http://www.sciencedirect.com/science/article/pii/S0306261917314915}

\bibitem{Castellanos2017}
J.A.F. Castellanos, D.~Coll-Mayor, J.A. Notholt, in \emph{2017 IEEE
  International Conference on Smart Energy Grid Engineering (SEGE)} (2017), pp.
  367--372.
\newblock \doi{10.1109/SEGE.2017.8052827}

\bibitem{Imbault2017}
F.~Imbault, M.~Swiatek, R.~de~Beaufort, R.~Plana, in \emph{2017 IEEE
  International Conference on Environment and Electrical Engineering and 2017
  IEEE Industrial and Commercial Power Systems Europe (EEEIC / I CPS Europe)}
  (2017), pp. 1--5.
\newblock \doi{10.1109/EEEIC.2017.7977613}

\bibitem{GuaranteesofOriginOfgem}
{Ofgem}.
\newblock Guarantees of origin {(GoOs)}.
\newblock
  \urlprefix\url{https://www.ofgem.gov.uk/environmental-programmes/rego/energy-suppliers/guarantees-origin-goos}

\bibitem{Murkin2018}
J.~Murkin, R.~Chitchyan, D.~Ferguson, in \emph{From Science to Society}, ed. by
  B.~Otjacques, P.~Hitzelberger, S.~Naumann, V.~Wohlgemuth (Springer
  International Publishing, 2018), pp. 139--151

\bibitem{Kounelis2017}
I.~Kounelis, G.~Steri, R.~Giuliani, D.~Geneiatakis, R.~Neisse, I.~Nai-Fovino,
  in \emph{2017 International Conference in Energy and Sustainability in Small
  Developing Economies (ES2DE)} (2017), pp. 1--6.
\newblock \doi{10.1109/ES2DE.2017.8015343}

\bibitem{Hahn2017}
A.~Hahn, R.~Singh, C.C. Liu, S.~Chen, in \emph{2017 IEEE Power Energy Society
  Innovative Smart Grid Technologies Conference (ISGT)} (2017), pp. 1--5.
\newblock \doi{10.1109/ISGT.2017.8086092}

\bibitem{Munsing2017}
E.~Münsing, J.~Mather, S.~Moura, in \emph{2017 IEEE Conference on Control
  Technology and Applications (CCTA)} (2017), pp. 2164--2171.
\newblock \doi{10.1109/CCTA.2017.8062773}

\bibitem{Mengelkamp2017}
E.~Mengelkamp, B.~Notheisen, C.N. Beer, D.~Dauer, C.~Weinhardt, Computer
  Science - Research and Development pp. 1--8 (2017)

\bibitem{Aitzhan2016}
N.Z. Aitzhan, D.~Svetinovic, IEEE Transactions on Dependable and Secure
  Computing \textbf{PP}(99), 1 (2016).
\newblock \doi{10.1109/TDSC.2016.2616861}

\bibitem{Scanergy}
Scanergy: A scalable and modular system for energy trading between prosumers.
\newblock \urlprefix\url{http://scanergy-project.eu/}

\bibitem{MihaylovJARMABG15}
M.~Mihaylov, S.~Jurado, N.~Avellana, I.S. Razo{-}Zapata, K.V. Moffaert,
  L.~Arco, M.~Bezunartea, I.~Grau, A.~Ca{\~{n}}adas, A.~Now{\'{e}}, in
  \emph{Proceedings of the 2015 International Conference on Autonomous Agents
  and Multiagent Systems, {AAMAS} 2015, Istanbul, Turkey, May 4-8, 2015}
  (2015), pp. 1917--1918

\bibitem{MihaylovJMAN14}
M.~Mihaylov, S.~Jurado, K.V. Moffaert, N.~Avellana, A.~Now{\'{e}}, in
  \emph{{SMARTGREENS} 2014 - Proceedings of the 3rd International Conference on
  Smart Grids and Green {IT} Systems, Barcelona, Spain, 3-4 April, 2014}
  (2014), pp. 101--106.
\newblock \doi{10.5220/0004960201010106}

\bibitem{MihaylovJurado2014}
M.~Mihaylov, S.~Jurado, N.~Avellana, K.V. Moffaert, I.M. de~Abril, A.~Nowé, in
  \emph{11th International Conference on the European Energy Market (EEM14)}
  (2014), pp. 1--6.
\newblock \doi{10.1109/EEM.2014.6861213}

\bibitem{15Firms}
J.~Deign.
\newblock 15 firms leading the way on energy blockchain.
\newblock
  \urlprefix\url{https://www.greentechmedia.com/articles/read/leading-energy-blockchain-firms}

\bibitem{alliander}
Alliander.
\newblock \urlprefix\url{https://www.alliander.com/en}

\bibitem{bankymoon}
Bankymoon.
\newblock \urlprefix\url{http://bankymoon.co.za/}

\bibitem{conjoule}
Conjoule.
\newblock \urlprefix\url{http://conjoule.de/en/home/}

\bibitem{drift}
Drift.
\newblock \urlprefix\url{https://www.joindrift.com/}

\bibitem{greeneum}
Greeneum.
\newblock \urlprefix\url{https://www.greeneum.net/}

\bibitem{gridplus}
Grid+.
\newblock \urlprefix\url{hhttps://gridplus.io/}

\bibitem{gridSingularity}
Grid singularity.
\newblock \urlprefix\url{http://gridsingularity.com}

\bibitem{electron}
Electron.
\newblock \urlprefix\url{http://www.electron.org.uk/}

\bibitem{LO3Energy}
Lo3 energy.
\newblock \urlprefix\url{hhttps://lo3energy.com/}

\bibitem{myBit}
Mybit.
\newblock \urlprefix\url{https://mybit.io/}

\bibitem{ponton}
Ponton.
\newblock \urlprefix\url{http://www.ponton-consulting.de/index.php}

\bibitem{powerLedger}
Powerledger.
\newblock \urlprefix\url{https://powerledger.io/}

\bibitem{solarCoing}
Solarcoin.
\newblock \urlprefix\url{https://solarcoin.org/}

\bibitem{sunExchange}
Sun exchange.
\newblock \urlprefix\url{https://thesunexchange.com/}

\bibitem{veridium}
Veridium labs.
\newblock \urlprefix\url{http://veridium.io/}

\bibitem{wePower}
wepower.
\newblock \urlprefix\url{https://wepower.network/}

\bibitem{Brooklyn2016}
A.~Rutkin.
\newblock Blockchain-based microgrid gives power to consumers in new york
  (2016).
\newblock \urlprefix\url{https://www.newscientist.com/article/2079334}

\bibitem{BrooklynGrid}
Brooklyn mircorgrid.
\newblock \urlprefix\url{https://www.brooklyn.energy/}

\bibitem{Exergy}
Lo3Energy.
\newblock Blockchain-based microgrid gives power to consumers in new york.
\newblock \urlprefix\url{http://lo3energy.com/}

\bibitem{enerchain}
Enerchain project.
\newblock \urlprefix\url{https://enerchain.ponton.de/}

\bibitem{frenchLaw2017}
France adopts law for self-consumption of renewable energy.
\newblock
  \urlprefix\url{https://www.pv-tech.org/news/france-adopts-law-for-self-consumption-of-renewable-energy}

\bibitem{australianReport2017}
AGL.
\newblock Peer-to-peer distributed ledger technology assessment.
\newblock
  \urlprefix\url{https://arena.gov.au/assets/2017/10/Final-Report-MHC-AGL-IBM-P2P-DLT.pdf}

\end{thebibliography}

% % Non-BibTeX users please use
% \begin{thebibliography}{}
% %
% % and use \bibitem to create references. Consult the Instructions
% % for authors for reference list style.
% % Format for Journal Reference
% Author, Article title, Journal, Volume, page numbers (year)
% Author, Book title, page numbers. Publisher, place (year)
% \end{thebibliography}

\end{document}